\documentclass[12pt]{iopart}

\usepackage{iopams}

\usepackage[mathscr]{eucal}
\usepackage{array}
\usepackage{epsfig}
\usepackage{psfrag} 
\usepackage{subfigure}

\usepackage{graphicx}
\usepackage{bm}

\usepackage{color}
\usepackage[normalem]{ulem} 

\newcommand{\Replace}[2]{\bgroup\noindent\textcolor{red}{\xout{#1} #2}\egroup\ignorespacesafterend}
\newcommand{\Delete} [1]{\bgroup\noindent\textcolor{red}{\xout{#1}}\egroup\ignorespacesafterend}
\newcommand{\Insert} [1]{\bgroup\noindent\textcolor{blue}{#1}\egroup\ignorespacesafterend}
\newcommand{\Comment}[1]{\definecolor{Mygray}{gray}{0.50}\bgroup\color{Mygray}\noindent#1\egroup\ignorespacesafterend}

\newcommand \Stefan [1] {\bgroup\noindent[\textcolor{blue}{\textbf{Stefan}: #1}]\egroup\ignorespacesafterend}
\newcommand \Michael[1] {\bgroup\noindent[\textcolor{blue}{\textbf{Michael}: #1}]\egroup\ignorespacesafterend}

\usepackage{pifont}              

\newcommand{\Bl}{{\boldsymbol{\mathnormal l}}}
\newcommand{\Bn}{{\boldsymbol{\mathnormal n}}}

\newcommand{\Br}{{\boldsymbol{\mathnormal r}}}
\newcommand{\BA}{{\boldsymbol{\mathnormal A}}}

\newcommand{\BQ}{{\boldsymbol{\mathnormal Q}}}

\newcommand{\superscr}[1]{\ensuremath{{}^{\rm #1}}}

\newcommand{\tauext }{\ensuremath{           \tau \superscr{ext}}}
\newcommand{\taub   }{\ensuremath{           \tau \superscr{b}}}
\newcommand{\tauy   }{\ensuremath{           \tau \superscr{y}}}
\newcommand{\tault  }{\ensuremath{           \tau \superscr{lt}}}

\newcommand{\Bkappa   }{\ensuremath{\boldsymbol\kappa}}

\newcommand{\rhot   }{\ensuremath{           \rho \superscr{t}}}
\newcommand{\qt   }{\ensuremath{                q \superscr{t}}}

\renewcommand{\div}{\textrm{div}}

\newcommand{\curl}{\textrm{curl}}

\newcommand{\eqref}[1]{(\ref{#1})}

\begin{document}

\title[Scaling Properties of Dislocation Simulations]{Scaling Properties of Dislocation Simulations in the Similitude Regime}

\author{Michael Zaiser and Stefan Sandfeld}
\address{Institute for Materials Simulation, Friedrich-Alexander University Erlangen-N\"urnberg (FAU), Dr. Mack-Strasse 77, 90762 F\"urth, Germany}
\ead{michael.zaiser@ww.uni-erlangen.de}

\date{\today}


\begin{abstract}
Dislocation systems exhibit well known scaling properties such as the Taylor relationship between flow stress and dislocation density, and the "law of similitude" linking the flow stress to the characteristic wavelength of dislocation patterns. Here we discuss the origin of these properties, which can be related to generic invariance properties of the equations of evolution of discrete dislocation systems, and their implications for a wide class of models of dislocation microstructure evolution. We demonstrate that under certain conditions dislocation simulations carried out at different stress, dislocation density, and strain rate can be considered as equivalent, and we study the range of deformation conditions ("similitude regime") over which this equivalence can be expected to hold. In addition, we discuss restrictions imposed by the stated invariance properties for density-based, non-local or stochastic models of dislocation microstructure evolution, and for dislocation patterns and size effects. 
\end{abstract}

\maketitle

\section{\label{intro} Introduction}

Dislocation patterns emerging and evolving under plastic deformation are often characterized by scaling relations which in the literature are commonly addressed as "law of similitude" \cite{Kuhlmann1962} or "similitude principle" and can be observed in a wide variety of materials and under various deformation conditions \cite{Raj1986,Sauzay11}. This principle states that the characteristic wavelength $d$ of dislocation patterns (cell size, spacing of dislocation walls) that form during deformation at a stress $\tau_{\rm f}$ is inversely proportional to the stress: $d = CGb/\tau_{\rm f}$, where $b$ is the length of the dislocation Burgers vector, $G$ is the shear modulus of the material, and the proportionality constant $C$ is approximately independent of material and deformation conditions. 

In conjunction with the even more general Taylor relationship $\tau_{\rm f} = \alpha G b \sqrt{\rho}$ which relates the flow stress $\tau_{\rm f}$ to the dislocation density $\rho$ through a non-dimensional constant $\alpha \approx 0.3$ \cite{Sauzay11}, the similitude principle can be re-phrased as $d = D \rho^{-1/2}$: The dislocation pattern wavelength is proportional to the average dislocation spacing, with a proportionality constant $D = C/\alpha$ which is typically of the order of 10 \cite{Raj1986}. In this form, the similitude principle has been demonstrated to hold over 4 decades in cell size and 8 orders of magnitude in dislocation density \cite{Rudolph05}. 

In view of the ubiquity of "similitude"-type behavior, it is natural to ask where lies the origin of this type of scaling behavior and what are its implications for dislocation simulations. While simulating the evolution of dislocation patterns at large strains is still beyond the capacity of present-day discrete dislocation dynamics models, such models are now quite capable of simulating the incipient formation of dislocation patterns at moderate strains and dislocation densities where the "similitude principle" generally holds \cite{Gomez06}. It is therefore important to ask which fundamental properties of dislocation systems are responsible for the formation of patterns with the observed scaling properties and what are the implications of these properties for simulations. We will approach this question from a somewhat unusual perspective: Rather than investigating what specific dislocation mechanisms are responsible for patterning, we will discuss generic invariance properties of the dynamic equations solved in both two- and three-dimensional dislocation simulations, and we will demonstrate that these invariance properties imply the similitude principle. We will then investigate  how mechanisms such as dislocation cross slip and dislocation annihilation may lead to deviations from  similitude, and we will determine the range of parameters where similitude scaling is expected to hold. We will apply our discussion of invariance properties to demonstrate how dislocation simulations performed at different density, stress, and strain rate can be related to each other, and to elucidate the minimal features of dislocation interactions which are needed in models or simulations in order to obtain dislocation patterns consistent with the similitude principle. We note that the scaling relations demonstrated in the following have been previously discussed by one of the present authors for the special case of two-dimensional dislocation systems evolving in single slip under a linear stress-velocity law \cite{Zaiser01,Zaiser02}. A comprehensive discussion of the mathematical background of "similitude scaling" has however, to our knowledge, never been published despite the simplicity of the underlying mathematical relations which appear to us almost self evident. 

\section{\label{Basic invariance theorem} The fundamental invariance theorem}

\subsection{Formulation for 2D dislocation systems}

We consider a generic 2D dislocation simulation where $N$ edge dislocations of different slip systems are represented by points in the $xy$ plane. The dislocations are either contained within an infinite crystal (the boundaries are remote such that image stresses can be neglected), or the system is replicated periodically. The Burgers vector of the $i$-th dislocation is $\vec{b}^{\,i} = s^{\,i} b \vec{e}^{\;i}$ where $s^{\,i}$ is the sign of the dislocation, $b$ is the Burgers vector modulus, and $\vec{e}^{\;i}$ is the unit vector in the slip direction which coincides with the dislocation glide direction. The slip plane normal of this dislocation is $\vec{n}^{\,i} = \vec{e}^{\;i} \times \vec{e}_z$. The driving force for the motion of the $i$th dislocation is given by the glide component of the Peach-Koehler force which is proportional to the resolved shear stress at the dislocation position:
\begin{equation}
\label{eq:tau}
\tau^{i} (\vec{r}^{\;i}) = \tau^{{\rm ext}, i} + \tau^{{\rm int},i}(\vec{r}^{\;i}) \;,
\end{equation}
where the "external" resolved shear stress is $\tau^{{\rm ext},i} = M^{i}_{kl} {\sigma}^{\rm ext}_{kl}$ and we apply the Einstein summation convention for lower indices. (Upper indices, instead, are understood as dislocation labels). $\sigma^{\rm ext}_{kl}$ is the externally applied stress field caused by displacements and/or tractions applied to the remote boundaries of the system, and the projection tensor $M^{i}_{kl}$ is defined as  $M^{i}_{kl} = (n^{i}_k e^{i}_l + e^{i}_k n^{i}_l)/2$ where $e^i_k$ and $n^i_k$ are the respective components of $\vec{e}^{\;i}$ and $\vec{n}^{\,i}$. Dislocation interactions are described by the internal stress field
\begin{equation}
\label{eq:taui}
\tau^{{\rm int},i}(\vec{r}^{\;i}) = M^{i}_{kl}  \sum_{j \neq i} \sigma_{kl}(\vec{b}^{\,j}, \vec{r}^{\;i}-\vec{r}^{\;j})\;,
\end{equation}
where $\sigma_{kl}(\vec{b}^{j}, \vec{r}^{\;i}-\vec{r}^{\;j})$ is the stress caused at $\vec{r}^{\;i}$ by a dislocation of Burgers vector $\vec{b}^{\;j}$ located at $\vec{r}^{\;j}$. We approximate the dislocation stress field by the stress field of a dislocation in an infinite body, in case of periodic boundary conditions complemented by that of its periodic images. 

The dislocation velocity is assumed to be a power of the driving force. Hence, the equations of motion of the dislocations are given by 
\begin{equation}
\label{eq:motion}
\frac{\partial \vec{r}^{\;i}}{\partial t} = \vec{v}^{\;i}(\vec{r}^{\;i})\quad,\quad
\vec{v}^{\;i} (\vec{r}^{\;i}) = v_0 \vec{e}^{\;i}  {\rm sign}(\tau^i) \left|\frac{\tau^i(\vec{r}^{\;i})}{G}\right|^n\;,
\end{equation}
where $v_0$ is a characteristic velocity and $n$ is the stress exponent. Two moving dislocations may react upon contact, forming either mobile dislocations of a third slip system or immobile barriers. No specific reaction radius is assigned to these reactions which, owing to the singularity of the stress fields, occur in finite time (this may pose practical problems from a numerical point of view but these are irrelevant as far as the mathematical structure of the equations is concerned).

{\bf Theorem:} The system of equations \eqref{eq:tau} and \eqref{eq:taui} is invariant under the transformation
\begin{equation}
\label{eq:trans}
\vec{r}_i \to \lambda \vec{r}_i\quad,\quad
\sigma^{\rm ext}_{kl} \to \lambda^{-1} \sigma^{\rm ext}_{kl}\quad,\quad
t \to \lambda^{n+1}t\quad.
\end{equation}

{\bf Proof:} Proof is obtained by substituting \eqref{eq:trans} into \eqref{eq:taui} and observing that the interaction stresses scale like $1/|\vec{r}^{\;i}-\vec{r}^{\;j}|$. Therefore, the dislocation stresses, \eqref{eq:taui}, decrease similarly to the external stress in proportion with $\lambda^{-1}$. As a consequence, in \eqref{eq:motion} upon insertion of \eqref{eq:trans} all factors $\lambda$ cancel. 

{\bf Corollary 1:} Suppose we have a solution  $\{\vec{r}^{\;i}(t)\}$ of the system (1,2) at some stress $\sigma_{kl}^{\rm ext}$. Let $\epsilon_{kl}^p(t)$ be the plastic strain which has accumulated due to dislocation motion during the evolution of this system from its initial state. Then there exists a solution $\{\vec{\tilde{r}}^{\;i}(\tilde{t})\}$ at the stress $\tilde{\sigma}_{kl}^{\rm ext} = \sigma_{kl}^{\rm ext}/\lambda$ where $\vec{\tilde{r}}^{\;i}(\tilde{t}) = \lambda \vec{r}^{\;i}$ and $\tilde{t} = \lambda^{n+1}t$. The plastic strain associated with this solution is $\epsilon_{kl}^p(\tilde{t}) = \epsilon_{kl}^p(t)/\lambda$.

{\bf Corollary 2:} For any stationary solution $\{\vec{r}^{\;i}\}$ of the system of equations (1,2), which is in static equilibrium at the stress $\sigma_{kl}^{\rm ext}$, there exists a one-parameter family of solutions $\{\lambda \vec{r}^{\;i}\}$ which are in static equilibrium at the respective stresses $\lambda^{-1}\sigma_{kl}^{\rm ext}$. If we associate, through some averaging procedure, a dislocation density $\rho$ to the solution $\{\vec{r}^{\;i}\}$, then these solutions are associated with the respective dislocation densities $\lambda^{-2}\rho$.

{\bf Corollary 3a:} If, in a dislocation simulation carried out at the stress $\sigma_{kl}^{\rm ext}$, a quasi-stationary pattern of wavelength $d$ and dislocation density $\rho$ emerges, then there exists a one-parameter family of stretched quasi-stationary patterns of wavelength $\lambda d$ and density $\rho/\lambda^2$ which emerge in simulations carried out at the stresses $\sigma_{kl}^{\rm ext}/\lambda$. These one-parameter families of solutions obey both the Taylor relationship and the "law of similitude".  

{\bf Corollary 3b:} If in a dislocation simulation a dislocation pattern of wavelength $d$ and dislocation density $\rho$ emerges which is metastable after unloading, then it belongs to a one-parameter family of patterns of wavelength $\lambda d$ and density $\rho/\lambda^2$ which are also metastable in the unloaded state. 

{\bf Remark 1:}
The above stated principle can be extended to nonlinear velocity laws describing thermally activated dislocation motion where the velocity law has the form 
\begin{equation}
\vec{v}^{\;i} (\vec{r}^{\;i}) = v_0 {\rm sign}(\tau^i) f \left(T,\frac{\tau^i}{S}\right)
\label{eq:thermalact}
\end{equation}
{\em if} the strain-rate sensitivity $S$ in this law is proportional to the dislocation spacing (in physical terms, if we are dealing with thermally assisted overcoming of obstacles such as forest dislocations. This is generally expected to be the velocity-controlling mechanism in pure fcc metals where the Cottrell-Stokes relation \cite{Cottrell55} holds. In Eq. (\ref{eq:thermalact}), the function $f(T,\tau)$ may be strongly non-linear, e.g. $f = \exp(- [(G_0 - \tau V_{\rm a})/(k_{\rm B}T)]$ where $k_{\rm B}$ is Boltzmann's constant and the activation volume $V_{\rm a}$ relates to the strain-rate sensitivity $S$ via $S = k_{\rm B}T/V_{\rm a}$. The system of equations (\ref{eq:motion}) with the velocity law (\ref{eq:thermalact}) is invariant under the transformation
\begin{equation}
\label{eq:trans1}
\vec{r}_i \to \lambda \vec{r}_i\quad,\quad
\sigma^{\rm ext}_{kl} \to \lambda^{-1} \sigma^{\rm ext}_{kl}\quad,\quad
t \to t\quad.
\end{equation}
and the above corollaries apply accordingly. 

{\bf Remark 2:} 

Irrespective of the velocity law, the invariance property \eqref{eq:trans} holds for static dislocation arrangements where all dislocations are stress free and which therefore correspond to metastable minima of the elastic energy. This observation must, however, not lead to the conclusion that quasi-static metastable dislocation patterns -- which include all patterns which can be observed ex situ -- universally obey similitude. Such a conclusion would be warranted only if the sequence of metastable minima reached, under an given loading path, by a dislocation system could be considered unique, i.e., independent of the dynamics of evolution in between these configurations. Such uniqueness is found in the motion of elastic manifolds or charge-density waves in random media where it is implicit in Middleton's 'no passing' theorem \cite{Middleton92} and the dynamics of particular dislocation configurations such as pile ups \cite{Moretti04} exhibits analogous properties. For general dislocation systems in two and three dimensions, however, a proof along the lines of Middleton is not possible because of the strongly anisotropoic nature of dislocation interactions which change their sign depending on the relative orientation of two dislocations (dislocation segments). As a consequence, the stated invariance principles can in general {\em not} be expected to hold in situations where the dislocation velocity is controlled by obstacles other than dislocations, such as Peierls barriers, radiation debris, solute atoms, precipitates etc. 

\subsection{Formulation for 3D dislocation systems}

The generalization of the considerations of the previous section towards 3D systems of dislocations requires some additional definitions and notations. We consider a situation where $N$ dislocation lines (numbered $i=1\ldots N$) with Burgers vectors $\vec b^{\,i}$ of modulus $b$ are positioned in slip planes ${\rm SP}_i$ with normal vectors $\vec n^{\;i}$. The dislocations initially form closed loops $C^i$ contained each within a single slip plane. These loops are labeled by $i = 1 \to N$ and  parameterized by $\vec{r}(s^i)$ with local tangent vector $\vec{t}(s^i) = {\rm d}\vec{r}/{\rm d} s^i$. In analogy to the previous section, we again assume that the dislocation loops are either contained within a quasi-infinite crystal where the boundaries are remote such that image stresses can be neglected, or the system is replicated periodically. 

Using the result that the displacement field of a closed planar dislocation loop can be derived using a Green's function together with the equilibrium conditions of elasticity theory \cite{deWit60}, the internal stress at a general position $\vec r$ can be written as a sum of line integrals over the loops:
\begin{eqnarray}
\sigma_{kl}^{\rm int}(\vec{r}) = -\frac{\mu}{8\pi} \sum_i \oint_{C^i} && \left\{  \frac{2}{1-\nu}\left( \frac{\partial^3 R}{\partial r_n \partial r_k \partial r_l} - \delta_{kl} \frac{\partial}{\partial r_n}\nabla^2R \right) b_o\epsilon_{nom}t_m \right. \nonumber\\
             &+& \left.\left(\frac{\partial}{\partial r_n} \nabla^2 R\right) b_o  \left[\epsilon_{nok}\, t_l    + \epsilon_{nol}\, t_k \right] \right\}{\rm d}s^i, 
          \label{eq:3Dloopstress}
\end{eqnarray}
where $\epsilon$ is the permutation symbol, $r_k$ are the components of $\vec r$, furthermore $R^i:=|\vec r(s^i) - \vec r|$. 
As the loops move, intersecting loops may react to form a dislocation network. In that case, the stress can still be evaluated according to \eqref{eq:3Dloopstress}, as the segment $\Sigma_i$ that results from a reaction can, from the point of view of stress calculations, be envisaged as a superposition of segments of the two intersecting loops. (This includes the case where dislocation annihilate through a collinear reaction \cite{Madec03} where $\Sigma$ is the superposition of two screw dislocations where the tangent vectors $\vec{t}$ have opposite sign such that the stress contributions cancel.) Alternatively to \eqref{eq:3Dloopstress}, we can then express the internal stress as a sum over all segments $\Sigma_j$ that either form closed loops or terminate at nodes where the sum of Burgers vectors is zero: 
\begin{eqnarray}
\sigma_{kl}^{\rm int}(\vec{r}) = -\frac{\mu}{8\pi} \sum_i \int_{\Sigma_j} && \left\{  \frac{2}{1-\nu}\left( \frac{\partial^3 R}{\partial r_n \partial r_k \partial r_l} - \delta_{kl} \frac{\partial}{\partial r_n}\nabla^2R \right) b_o\epsilon_{nom}t_m \right. \nonumber\\
             &+& \left.\left(\frac{\partial}{\partial r_n} \nabla^2 R\right) b_o  \left[\epsilon_{nok}\, t_l    + \epsilon_{nol}\, t_k \right] \right\}{\rm d}s^i, 
          \label{eq:3Dstress}
\end{eqnarray}

The resolved shear stress is obtained from the stress tensor as in the 2D case, 
\begin{equation} 
\tau(s^i) = M_{kl}\left( {\sigma}^{\rm ext}_{kl} + \sigma^{\rm int}_{kl}(\vec{r}(s^i))\right).
\end{equation}
The integral over $C^i$ in Eq. (\ref{eq:3Dstress}) has in this case to be understood as the principal value. 
The dislocation glide velocity is, in analogy to the 2D case, assumed to be given by
\begin{equation} \label{eq:3D:v}
\frac{\partial \vec{r}(s^i)}{\partial t} = \vec{v}(s^i) = 
v_0(\vec{t}) \vec{e}(s^i) {\rm sign}(\tau(s^i)) \left|\frac{\tau(s^i)}{G}\right|^n\;,
\end{equation}
where the local glide direction $\vec{e}(s^i)$ is given by $\vec{e}(s^i) = \vec{t}(s^i) \times \vec{n}^{\;i}$ and the characteristic velocity $v_0$ depends in general on the local orientation of the dislocation line.

{\bf Theorem 1A:} The system of equations (\ref{eq:3Dstress})-(\ref{eq:3D:v}) is invariant under the transformation (\ref{eq:trans}). Corollaries 1-3a/b and Remarks 1,2 hold accordingly also for the 3D case. 
 
{\bf Proof:} Proof is obtained by substituting \eqref{eq:trans} into \eqref{eq:3Dstress} and noting that the differential dislocation line length rescales like ${\rm d}s^i \to \lambda {\rm d}s^i$. One then sees that the internal stresses decrease again in proportion with $\lambda^{-1}$. As a consequence, in \eqref{eq:3D:v} again all factors $\lambda$ cancel. 

Thus, we observe that exactly the same invariance principle applies to 2D and to 3D dislocation simulations. Of course, there is one important difference between the described simulation settings, since the described 2D simulation setting does not consider multiplication, whereas the 3D setting naturally incorporates changes in dislocation line length, e.g. line length increases because of loop expansion. This leads to an important conclusion regarding models which introduce dislocation multiplication into 2D simulations by way of phenomenological rules: If such models aim at being consistent with 3D dislocation dynamics, the multiplication rules must be constructed in such a manner that the resulting dynamics still fulfills Theorem 1 stated above. This happens naturally when multiplication rates are directly matched to 3D simulations  \cite{Gomez06}, whereas other rules need to be adapted carefully. For instance, the introduction of dipoles with "nucleation width" $L_{\rm nuc}$ (see e.g. \cite{VdG95}) is consistent with Theorem 1 only if the nucleation width is, in the course of a simulation, adapted to evolve in proportion with the mean dislocation spacing as new sources are being added. 

\subsection{Line tension approximation}

An approximate method for treating the self-interaction of curved dislocation lines is to use a line tension approximation. In this case one replaces the internal stress field by a local term that is proportional to the line curvature, i.e., one writes
\begin{equation} 
\tau(s) = M_{kl} {\sigma}^{\rm ext}_{kl} + T(\theta)/R(s) \;,
\end{equation}
where $T(\theta)$ is a dislocation line tension (energy per unit line length) where $\theta$ characterizes the dislocation line character, and $R = |\partial^2 \vec{r}/\partial s^2|$ is the local curvature radius of the dislocation line. If we assume $T(\alpha) \propto Gb^2$ to be independent of the dislocation arrangement, it is immediately evident that, upon the transformation \eqref{eq:trans} the curvature radii multiply with $\lambda$ and thus the self-interaction stresses again decrease like $\lambda^{-1}$. Hence, treating dislocation interactions in this approximation maintains the basic transformation invariance of 3D dislocation systems. Patterns forming in models which use a line tension approximation (see e.g. \cite{Gomez06}) are therefore bound to fulfill the similitude principle. 

However, the dislocation line energy is not strictly independent on dislocation density: For a screened dislocation arrangement it exhibits a logarithmic density dependence, $T \propto G b^2 \log(b\sqrt{\rho})$. Including this dependence implies corrections to similitude, such as a logarithmic dislocation density dependence of the pre-factor $\alpha$ in the Taylor relationship as observed in experiment \cite{Basinski79} and commonly explained in terms of the line tension model \cite{Basinski79,Madec02}. This issue will be addressed in Section \ref{Limits} in more detail. 

\section{\label{stresstraincontrol} Stress and strain rate controlled simulations}

Until now our considerations have been based upon the assumption of a constant stress. We now ask how these need to be modified if (i) stress is ramped up at a constant rate, or (ii) if stress is related to an imposed strain rate through a "machine equation". In these cases \eqref{eq:tau} and \eqref{eq:taui} need to be supplemented by an equation for the stress evolution. In case of stress controlled testing with constant stress rate this can be written as
\begin{equation}
\frac{\partial {\bm \sigma}_{\rm ext}}{\partial t} = \dot{\Theta} {\bm \Sigma} 
\end{equation}
where $\dot{\Theta}$ is the characteristic stress rate and ${\bm \Sigma} = {\bm \sigma}_{\rm ext}/{\sigma_{\rm eq}}$ is a non-dimensional tensor which may be written as the ratio of the stress tensor and the corresponding equivalent stress $\sigma_{\rm eq} = \sqrt{(3/2){\bm \sigma}_{\rm ext}':{\bm \sigma}_{\rm ext}'}$. This equation is invariant upon the rescaling \eqref{eq:trans1} if we simultaneously re-scale the initial stress according to \eqref{eq:trans1} and the stress rate according to 
\begin{equation}
\dot{\Theta} \to \lambda^{-n-2} \dot{\Theta} \;.
\end{equation}
In case of strain controlled testing, the external stress is imposed through displacements acting on the remote boundaries of the system which create a homogeneous stress state over the simulated volume. This can be expressed as
\begin{equation}
\frac{\partial {\bm \sigma}_{\rm ext}}{\partial t} = {\bm C \normalfont}\left[\dot{\bm \varepsilon}_{\rm ext} - \frac{1}{V} \int_V \dot{\bm \varepsilon}_{\rm pl}(\vec{r})  d^D r\right]
\label{eq:straincontrol}
\end{equation}
where $\bm C$ is the tensor of elastic constants, $\dot{\bm \varepsilon}_{\rm ext}$ is the remotely imposed strain rate, $V$ is the system volume, and $D=2,3$ for 2D and 3D dislocation systems, respectively. The local strain rate tensor $\dot{\bm \varepsilon}_{\rm pl}(\vec{r})$ is for 2D dislocation systems given by
\begin{equation}
\dot{\bm \varepsilon}_{\rm pl}(\vec{r})= \sum_i {\bm M}^i  b\, \vec{e}^{\,i} \vec{v}^{\;i} \delta(\vec{r}-\vec{r}^{\;i}), 
\label{eq:ep2D} 
\end{equation}
and for 3D systems, it is given by
\begin{equation}
\dot{\bm \varepsilon}_{\rm pl}(\vec{r}) = \sum_i {\bm M}^i \int_{C^i} b\, \vec{e}(s^i) \vec{v}(s^i)  \delta(\vec{r} - \vec{r}(s^i)) {\rm d}  s^i, 
\label{eq:ep3D}
\end{equation}
Under the rescaling \eqref{eq:trans1}, the domain $V$ of integration in \eqref{eq:straincontrol} transforms according to $V \to \lambda^D V$ while the dislocation velocities transform according to $\vec{v}_i \to \lambda^{-n} \vec{v}_i$, the Dirac function scales like $\lambda^{-D}$ and the line length in \eqref{eq:ep2D} increases like $s^i \to \lambda s^i$. Thus, both equations \eqref{eq:ep2D} and \eqref{eq:ep3D} are invariant under the transformation \eqref{eq:trans} if we simultaneously re-scale the imposed strain rate according to
\begin{equation}
\dot{\bm \varepsilon}_{\rm ext} \to \lambda^{-n-2} \dot{\bm \varepsilon}_{\rm ext}.
\end{equation}
In conclusion, under conditions of stress or strain rate control, simulation of a system of dislocation density $\rho$ and size $L$ at an imposed stress rate $\dot{\Theta}$ (strain rate $\dot{\bm \varepsilon}_{\rm ext}$) is equivalent to simulation of a system of dislocation density $\rho/\lambda^2$ and size $\lambda L$ at a stress rate  $\lambda^{-n-2} \dot{\Theta}$ (strain rate $\lambda^{-n-2} \dot{\bm \varepsilon}_{\rm ext}$). This observation holds for both 2D and 3D dislocation systems.  

\section{\label{Limits} Limits of similitude}

\subsection{Short-range self-interaction of curved dislocation lines}

An important technical problem in 3D dislocation dynamics simulations arises from the fact that the expression \eqref{eq:3Dloopstress} for the internal stress field becomes singular on the dislocation lines themselves. Whereas for a straight infinite dislocation the self force resulting from the singular stress terms is zero, the same is not true for general curved dislocations where the self force diverges. To regularize this divergence, various methods have been proposed in the literature \cite{Brown64,Gavazza76,Zbib98, Cai06}. All these methods have in common that they introduce an additional intrinsic length $\xi$ which physically relates to the extension of the dislocation core. As it would be unphysical to re-scale this length according to the stretching transformation $(\ref{eq:trans})$, this means that the self-interaction of a dislocation loop (dislocation segment) does not strictly obey similitude. (The interaction between different loops/segments, on the other hand, strictly obeys the stated invariance principles). The consequences are best illustrated by considering the self interaction in a line tension approximation. Cai et. al. \cite{Cai06} evaluate the line energy of a circular loop of radius $R$ (up to terms of the order of $(\xi/R)^2$) as
\begin{equation}
E(R)=2\pi R \frac{G b^2}{8\pi}\left(\frac{2-\nu}{1-\nu}\left[\ln\frac{8R}{\xi}-2\right]+\frac{1}{2}\right)\quad,
\end{equation}
with the average line tension
\begin{equation}\label{eq:T}
T = \frac{1}{2\pi} \frac{\partial E}{\partial R} = \frac{G^2}{8\pi}\left(\frac{2-\nu}{1-\nu}\left[\ln\frac{8R}{\xi}-1\right]+\frac{1}{2}\right)\quad.
\end{equation}
Thus, the line tension has a contribution which depends logarithmically on the ratio $R/\xi$ which leads to deviations from similitude scaling. To assess the order of magnitude of these deviations, let us consider the typical values $\nu = 1/3$, $\xi = 0.25$ nm, and compare two loops of radii $R$ and $R' = \lambda R$. The corresponding line tension values differ by a constant amount $\Delta T =  \frac{G^2}{8\pi}\frac{2-\nu}{1-\nu}\ln \lambda$. Assuming that the flow stress is exclusively controlled by line tension effects, $\sigma_{\rm f} \propto T/R \propto T\sqrt{\rho}$, we find that the ensuing relative correction to the flow stress is $\Delta \sigma/\sigma = \Delta T/T$. This is shown in Fig. 1 as a function of the dislocation density $\rho$ and the parameter $\lambda$, taking $R \approx \rho^{-1/2}$ and using the parameters $\nu = 0.3,\; \xi =0.25$\,nm. For illustration: With $\rho=10^{13}/{\rm m}^2$,  an increase in dislocation density by a factor $\lambda^2=64$ implies an increase in flow stress by a factor $\lambda=8$. The actual increase that follows from \eqref{eq:T} is about 25\% less (circle in Fig. 1). In typical dislocation simulations which cover the initial stages of deformation, the ensuing corrections to similitude scaling are small but not  negligible. The corresponding  corrections to the Taylor relationship are well known, see the review of Basinski and Basinski \cite{Basinski79}.

\begin{figure}
\begin{center}
\includegraphics[width=0.7\textwidth]{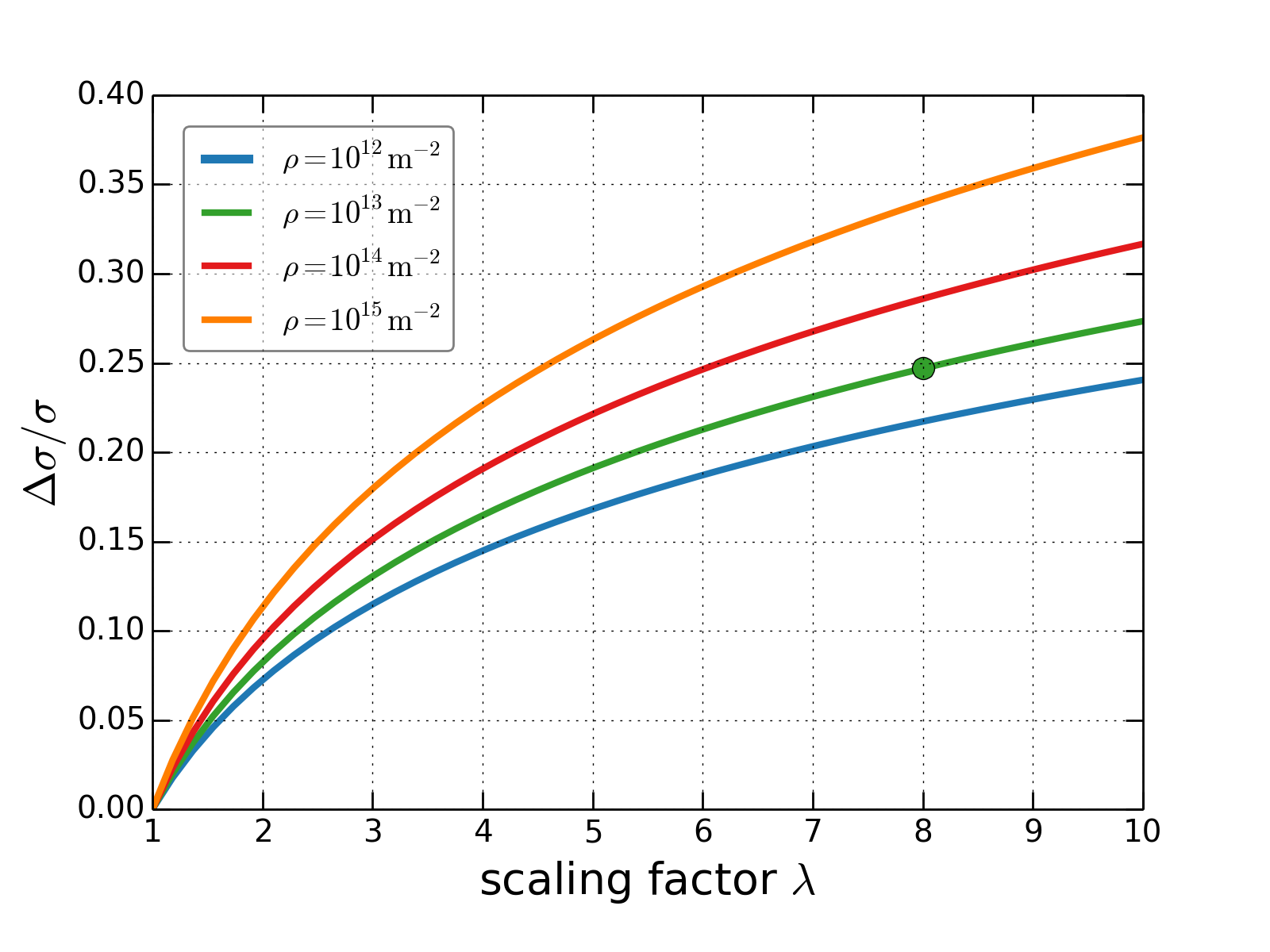}
\end{center}
\vspace*{-0.2cm}
\caption{Flow stress corrections  due to changes in line tension as a function of the scaling factor $\lambda$ for different reference dislocation densities $\rho$ (reference curvature radius $R \propto \rho^{-1/2}$)}
\label{fig:linetension}
\end{figure}

\subsection{Direct annihilation of non-screw dislocations}

Even in the absence of cross slip, dislocations of the same slip system may annihilate if the spacing of the respective slip planes falls below a critical value $y_{\rm e}$  \cite{Essmann79}. For dislocations of general orientation the atomic rearrangements occurring during this process lead to the formation of point defect agglomerates: We are dealing with a process related to the discreteness of the atomic lattice structure near the dislocation core. Again, $y_{\rm e}$ provides an additional length which does not rescale under the stretching transformation $(\ref{eq:trans})$ and therefore introduces corrections to similitude scaling. The conditions when this becomes relevant are straightforward to estimate: For dislocations of density $\rho$ on a given slip system, annihilation occurs after a mean free path $l_{\rm e} = 1/(\rho y_{\rm e})$. This corresponds to a critical shear strain $\gamma_{\rm e} = \rho b l_{\rm e} = b/y_{\rm e}$ which is independent of dislocation density. With $y_{\rm e} \approx 1.6$nm as suggested by Essmann for Cu \cite{Essmann79}, this strain is of the order of 20\%. As most current dislocation dynamics simulations extend only to much smaller strains, one does not expect to see much annihilation in these simulations. An exception are some published simulations where initial dislocation densities were extremely high, see e.g. \cite{Miguel02} where initial dislocation densities as high as $0.04 y_{\rm e}^{-2}$ were used. Such densities exceed, however, the values typically found in heavily deformed metals by one to two orders of magnitude. We also note that the removed dislocation configurations are extremely narrow dipoles which do not carry long-range stress fields and do not contribute to plastic flow. Hence, the impact of direct annihilation processes on the collective dynamics of dislocations is expected to be generally small. 

\subsection{Jog formation}

Cutting of forest dislocations creates jogs on dislocations. Jogged dislocation loops represent non-planar dislocation configurations which can in general not be decomposed into systems of planar loops. Thus, equations \eqref{eq:3Dloopstress} and \eqref{eq:3Dstress} are no longer strictly valid for such configurations. However, estimates based on typical dislocation glide paths of some tens of dislocation spacings and typical dislocation spacings of the order of 1000b show that the atomic density of jogs on dislocations arising from cutting processes and the ensuing corrections to the stress fields are small. This may be different in situations where climb processes have an appreciable influence on dislocation motion.  

\subsection{Cross slip and annihilation of screw dislocations}

Whether or not our considerations apply to materials where screw dislocation cross slip is prominent, depends on the factors which control the cross slip process. If cross slip is envisaged as an essentially athermal, stress controlled process as proposed by Brown \cite{Brown02} and recently explored by Paus and co-workers \cite{Paus13}, then our considerations of similitude scaling apply also to such processes. Cross slip of the screw part of a dislocation loop on a new slip plane creates a new, non-planar dislocation configuration. However, this process can be envisaged as the nucleation of a new loop on the cross slip plane (one segment of this loop cancels the originally cross slipped segment), and therefore \eqref{eq:3Dloopstress} and \eqref{eq:3Dstress} can still be used for evaluating the stress field of the resulting non-planar configuration. The same is true if the cross slip process leads to annihilation of two screw segments moving on different slip planes. Accordingly, all these processes must fulfill similitude scaling as discussed in previous sections. 

However, separate considerations apply if cross slip requires the overcoming of a stress-dependent energy barrier (formation of a constriction in a split dislocation) with the aid of thermal activation. This is the classical viewpoint on cross slip (see e.g. \cite{Thornton62}) which also underlies most implementations of cross slip in 3D DDD codes (see e.g. \cite{Kubin92,Rhee98}). Irrespective whether the parameters governing the cross slip process are taken from experiment (e.g. \cite{Bonneville88}) or atomistic simulation \cite{Bulatov98}, if controlled by thermal activation the rate of thermally activated cross slip will depend on stress and temperature in a strongly nonlinear (exponential) manner and therefore this process is expected to lead to characteristic deviations from similitude scaling. The same is true for other thermally activated processes which may influence the motion of dislocations, such as thermally assisted overcoming of Peierls barriers, solute atoms, or precipitates. Conversely, for processes where similitude scaling is strictly observed, we may expect that the above mentioned thermally activated processes are of secondary importance. 

\section{Discussion and Conclusions}

Our discussion of scaling relations in the dynamics of dislocation systems can be summarized into two simple statements: (i) If dislocation dynamics is mainly controlled by the elastic interactions between dislocations, then all characteristic lengths in evolving dislocation arrangements scale in proportion with the mean dislocation spacing (inverse square root of dislocation density). (ii) Under the same conditions, all characteristic stresses, in particular the flow stress, scale in proportion with the square root of dislocation density. These statements are supposed to hold whenever the dynamics of dislocations is mainly controlled by their elastic interactions, and processes on the scale of the dislocation cores (formation of constrictions in cross slip, overcoming of Peierls barriers, annihilation of parallel edge dislocations, presence of jogs), as well as interactions of dislocations and other defects (solute atoms, precipitates, radiation debris or grain boundaries) play a secondary role. As a paradigm, we may consider deformation of pure face-centered cubic metals at low to intermediate stresses/dislocation densities.  

Empirically, both findings have been known for decades ("law of similitude", "Taylor relationship") and most researchers working on dislocation simulation will be aware of them. However, the far-reaching consequences of these relations are not always realized: 

\begin{enumerate}
\item Implications for discrete simulations:
\begin{itemize}
\item The scale of a discrete dislocation simulation is not determined by the linear dimension $L$ of the simulated volume, but by the size in units of dislocation spacings. Thus in 2D, size is governed by $N = L^2 \rho$ (the number of dislocations), and in 3D, by the non-dimensional number $L^3 \rho^{3/2}$. 
\item The scale of strain in bulk dislocation simulations is given by $b \sqrt{\rho}$, and the scale of stress by $G b \sqrt{\rho}$. The time scale, as stated above, scales in proportion with the scale of stress to the power $-(n+1)$, or the scale of dislocation density to the power $-(n+1)/2$. These relations need to be kept in mind when comparing dislocation simulations with different parameters. To take an example: A 3D DDD simulation of a cube of edge length 1$\mu$m with an initial dislocation density of $10^{14}$ m$^{-2}$, a strain rate of $5000$ s$^{-1}$, and an end strain of 10\%, using a linear stress-velocity law, may be almost equivalent to simulation of a cube of edge length 10$\mu$m with an initial dislocation density of $10^{12}$ m$^{-2}$, a strain rate of $500$ s$^{-1}$, and an end strain of 1\%. 
\item If 3D processes such as dislocation multiplication or junction formation are incorporated into 2D dislocation dynamics models, care must be taken to make sure that the rules introduced are consistent with the scaling properties of bulk dislocation dynamics. For instance, a dislocation multiplication rule for 2D DDD simulations may be specified by requiring that a source of length $l_{\rm nucl}$ produces a dipole of the same width if, over a nucleation time $t_{\rm nucl}$, the resolved shear stress at the site of the source remains above a level $\tau_{\rm nucl} \propto Gb/l_{\rm nucl}$ \cite{VanderGiessen95,Benzerga04}. This rule is invariant under the transformation (\ref{eq:trans}) if the source length $l_{\rm nucl}$ is taken to be proportional to the dislocation spacing in the vicinity of the source and if the nucleation time scales in inverse proportion with the square of the externally applied shear stress $\tau_{\rm ext}$, or in proportion with $(\tau_{\rm ext}\tau_{\rm nucl})^{-1}$ as proposed in \cite{Benzerga04}. Thus, the scaling relations we have stated provide guidance whether 2D rules introduced to mimic 3D dislocation processes such as multiplication or junction formation are consistent with the properties of dislocation systems or not. If they are, then the resulting 2D dynamics will by construction obey the principle of similitude as indeed observed in corresponding simulations \cite{Gomez06}. 
\end{itemize}

\item Implications for density-based and statistical models 
\begin{itemize}
\item Attempts to model dislocation microstructure evolution in terms of the evolution of dislocation densities, irrespective of whether they consider deterministic \cite{Holt70,Ananthakrishna81,Walgraef85,Groma97,Groma03} or stochastic \cite{H"ahner96,H"ahner99}, space-dependent \cite{Holt70,Walgraef85,Groma97,Groma03} or space independent \cite{Ananthakrishna81,H"ahner96} evolution equations, should make sure that the formulated evolution equations are, at least in those cases where the underlying discrete dynamics is expected to obey similitude scaling, invariant under the transformations (\ref{eq:trans}) or (\ref{eq:trans1}). Equations which are not, cannot represent dislocations. This is a problem with many early models of nonlinear phenomena in dislocation systems (e.g. \cite{Ananthakrishna81,Walgraef85}) where parameters are not sufficiently well specified to decide whether or not they fulfill this basic requirement. 
\item
We can use the scaling relations (\ref{eq:trans}) or (\ref{eq:trans1}) as heuristic tools to identify the structure of admissible terms in evolution equations. For instance if, in the spirit of Aifantis \cite{Aifantis84} we want to introduce a second gradient of strain in the flow stress expression, then from the stated scaling principles (strain scales in proportion with $\rho^{1/2}$, stress scales in proportion with $G b \rho^{1/2}$), it follows that the concomitant internal length scale must scale in proportion with the dislocation spacing for the resulting term to be consistent with (\ref{eq:trans}) - as is indeed found when such terms are derived by averaging the underlying discrete dynamics of dislocations \cite{Groma03}. 
\end{itemize}
The analysis of density-based models for consistency with similitude scaling is exemplified in the Appendix where we consider four models - the dislocation patterning model formulated by Holt in 1970 \cite{Holt70}, a model of dynamic dislocation patterning during plastic flow, a stochastic model of fractal dislocation cell structure formation that was formulated by H\"ahner and Zaiser in the 1990s \cite{H"ahner99}, and the recently proposed 3D continuum dislocation dynamics model of Hochrainer and co-workers \cite{Hochrainer2014_JMPS}.

\item Implications for dislocation patterning, statistical properties of plastic flow, and size effects.

\begin{itemize}
\item In the regime of similitude scaling all characteristic lengths of bulk dislocation arrangements are bound to scale in proportion with the dislocation spacing. This is in particular true for the characteristic lengths of dislocation patterns - if such patterns form - and for the characteristic mean free path which governs dislocation storage and, hence, strain hardening. As a consequence of scaling invariance, meaningful statistical information about dislocation mean free paths and hardening can be extracted from discrete dislocation simulations, even if such simulations are "too small" (or confined to too low strains) to replicate the full phenomenology of dislocation patterning \cite{Devincre08}.
\item
As a consequence, any model (whether discrete or density based) which can reproduce dislocation patterning and which is consistent with dislocation properties must, in the regime where dislocation dynamics is controlled by elastic dislocation interactions, produce patterns with wavelengths that are proportional to the dislocation spacing, and inversely proportional to the stress at which they have formed. Conversely,  patterns that do not have these properties \cite{Zhou12} may be artefacts of initial conditions, boundary conditions or numerical errors. 
\item 
Any statistical signatures of plastic flow and dislocation dynamics that can be deduced from simulations or statistical models of elastically interacting dislocations must be consistent with the stated principles. Hence, if dislocation systems at some critical stress undergo a jamming or depinning transition, then it is from their scaling properties self-evident they must do so at any dislocation density -- with a critical stress that scales in proportion with the square root of that density \cite{Tsekenis11}. Even scale free power law distributions as observed for instance for strain bursts and slip avalanches \cite{Zaiser06} must comply with these scaling principles: Since strain is proportional to the total area swept by dislocations per unit volume, it is clear that the upper and/or lower limits of power law regimes in strain burst size distributions must scale in proportion with the square root of dislocation density. Similarly, if the spatial structure of slip avalanches can be associated with a characteristic correlation length, this length must be proportional to the dislocation spacing. If we consider scale free distributions of internal lengths (cell sizes), as in the model of fractal dislocation structures proposed by H\"ahner and Zaiser \cite{H"ahner99,Zaiser99}, then the upper and lower ends of the fractal scaling regime must scale in proportion with the mean dislocation spacing, and in inverse proportion with the flow stress \cite{Zaiser99}. 
\item
An interesting application of the stated principles concerns size effects. As long as no other length scales are relevant (e.g. the size of surface heterogeneities that control dislocation nucleation at the surface) a system of size $L$ and dislocation density $\rho$ at stress $\tau$ behaves similarly to a system of size $\lambda L$ and dislocation density $\rho/\lambda^2$ at stress $\tau/\lambda$. Let us assume that $\tau$ is a size dependent flow stress of the form $\tau = \tau_{\infty} [1 + \Delta(L,\rho)]$ where $\tau_{\infty}$ is the flow stress of the bulk system ($L \to \infty)$. It follows for the size dependent contribution $\Delta$ that for any $(\rho,\lambda)$ the relation $\Delta(L,\rho) = \Delta(\lambda L,\rho/\lambda^2)$ must be fulfilled. In other words, the size dependent fraction of the flow stress can depend only the product $(L \sqrt{\rho})$ but not on $L$ and $\rho$ separately. This has indeed been observed in simulations \cite{Tang07}:  larger systems with lower initial dislocation density behave similarly to smaller systems with higher dislocation density. It follows that comparisons of size dependent flow stresses without data for the corresponding dislocation microstructures, as commonly presented in the literature \cite{Greer11,Uchic09}, may be of limited usefulness. 

\end{itemize}

\end{enumerate}

In conclusion, we have stated invariance principles which are in a sense trivial. Nevertheless, they can serve as extremely useful tools to gauge the plausibility of simulation results, to compare simulations carried out on apparently different length and time scales, and to assess the validity of microstructure evolution models. In practice, in any given situation there can be many reasons why the stated principles may not or only partially apply - for instance, in dispersion- or precipitation hardened materials due to the length scales associated with the phase microstructure, in polycrystals due to the influence of grain boundaries, in low-temperature deformation of bcc metals due to the paramount influence of Peierls stresses, and in general in all situations where atomic-scale processes (dislocation nucleation/annihilation, cross slip etc.) are of importance. Nevertheless, these principles are at the core of the behavior of "pure" dislocation systems and provide a useful guideline for assessing the viability and performance of a broad class of models that purport to describe dislocation microstructure evolution.

\section*{Acknowledgment}

We gratefully acknowledge financial support from the Deutsche Forschungsgemeinschaft
(DFG) through Research Unit FOR1650 ‘Dislocation-based Plasticity (DFG grant No
SA2292/1-1) and the European m-era.net project ’FASS’ (grant No SA2292/2).

\section*{Appendix: Scaling analysis of some density-based dislocation models}

\section*{Appendix A: The patterning model of Holt}

\setcounter{equation}{0}
\renewcommand{\theequation}{A\arabic{equation}}

Holt \cite{Holt70} proposed in 1970 a model of dislocation patterning at zero applied stress, constructed in analogy with contemporary models of spinodal decomposition. He assumes the interaction energy functional for a system of screw dislocations of density $\rho$ in the form
\begin{equation}
E(\rho) = \int 2 \pi r f(r,\rho) \frac{Gb^2}{2 \pi} \ln \frac{R_0}{r} {\rm d}r \;,
\end{equation} 
where $f$ is a radially dependent pair correlation function describing the excess of screw dislocations of opposite sign surrounding a given dislocation. This function is normalized, $\int 2 \pi r f(r,\rho) {\rm d}r = 1$, and assumed by Holt in a phenomenological manner (we note that, for a system of edge dislocations, the corresponding function has much later been explicitly computed by Groma and co-workers \cite{Groma06}). The energy change associated with spatially dependent fluctuations $\delta \rho(\Br)$ around a homogeneous state $\rho_0$ follows in a long-wavelength approximation as 
\begin{equation}
\delta E(\rho) \approx - F_1 \delta \rho(\Br) - F_2 \Delta (\delta \rho(\Br))\;, 
\end{equation} 
where $\Delta$ is the Laplace operator and $F_1$ and $F_2$ are given by 
\begin{eqnarray}
F_1 &=& \frac{1}{\rho_0}  \int 2 \pi r f(r,\rho) \frac{Gb^2}{2 \pi} \ln \frac{R_0}{r} {\rm d}r\;,\nonumber\\
F_2 &=& \frac{1}{\rho_0}  \int \frac{\pi r^3}{2} f(r,\rho) \frac{Gb^2}{2 \pi} \ln \frac{R_0}{r} {\rm d}r\;.
\end{eqnarray} 
In the spirit of linear irreversible thermodynamics, the flux ${\bm j} = \rho B {\bm \nabla} (\delta E(\rho))$ of dislocations is assumed to be proportional to the gradient of the energy fluctuation, i.e., to the net force. Assuming that the dislocation density $\rho$ is a conserved quantity, it follows that
\begin{equation}
\frac{\partial \delta \rho}{\delta t} = - \rho_0 B \Delta[F_1 \delta \rho + F_2 \Delta (\delta \rho)]
\label{eq:deltarhoholt}
\end{equation}
or, in Fourier space
\begin{equation}
\frac{\partial \delta \rho(k)}{\delta t} =  \rho_0 B k^2 [F_1 - F_2 k^2] \delta \rho(k)\;.
\end{equation}
If both $F_1$ and $F_2$ are positive, then long-wavelength fluctuations are undamped with a dominant wavelength emerging at $\lambda = 2 \pi (2 F_2/F_1)^{1/2}$. To study the behavior of Holt's model under the scaling transformation (\ref{eq:trans}), we observe that the proportionality of the dislocation flux and the energy gradient (the force) implies that, for this model, the rate exponent has the value $n = 1$. Upon inserting the transformation (\ref{eq:trans}) into (\ref{eq:deltarhoholt}) we see that the model is invariant only if the parameters $F_1$ and $F_2$ transform according to $F_1 \to \lambda^2 F_1$, $F_2 \to \lambda^4 F_2$. Since the same transformation must also preserve the normalization of the pair correlation function $f$, this requires the pair correlation function $f$ to possess the structure $f(\rho,r) = \rho \phi(u)$ with 
$u = r \sqrt{\rho}$ and $\int 2 \pi u \phi(u) {\rm d} u = 1$. It then follows immediately that the dominant wavelength of the emergent dislocation pattern is proportional to $1/\sqrt{\rho}$, i.e., to the dislocation spacing. We thus observe that, if the pair correlation function in Holt's model is chosen in a manner that is consistent with the scaling  properties of dislocation systems, then the resulting dominant wavelength turns out to be consistent with similitude. Holt arrives in his 1970 paper \cite{Holt70} at the same result by way of several unnecessary ad-hoc assumptions -- e.g., he claims without proof that the correlation function should be proportional to $1/r$ and then introduces a finite-scale cut-off at a distance that is proportional to the dislocation spacing. None of these assumptions are necessary to arrive at the main result, which in fact derives from any normalized function of the structure $f(\rho,r) = \rho \phi(r \sqrt{\rho})$.

\section*{Appendix B: A model of dynamic dislocation patterning}

\setcounter{equation}{0}
\renewcommand{\theequation}{B\arabic{equation}}

Holt's model has been justly criticized for physical reasons. Dislocation patterns do not form close to thermal equilibrium, in absence of external stresses driving dislocation motion. Also, if one takes Holt's model at face value, then it is easy to see that for a system of parallel screw dislocations containing equal numbers of dislocations of both signs, the most efficient way to reduce the internal energy is to annihilate all the dislocations - a process formally excluded by Holt when he writes down the equation for the dislocation density as a conserved field. 

In the following we demonstrate that, by using a pair correlation function of the correct structure to describe dislocation interactions, it is easy to formulate physically more plausible models which produce spontaneous symmetry breaking and dislocation patterning consistent with the similitude principle. As an example, we consider a system of equal numbers of straight parallel edge dislocations of both signs on a single slip system, moving by glide under a constant external resolved shear stress $\tau_{\rm ext}$. The slip direction is taken to be the $x$ direction of a Cartesian coordinate system. Since dislocation motion is constrained to a set of parallel glide planes, no annihilation is considered (see Section 4.2 for a discussion of this point). The equations of motion for the densities $\rho^+$ and $\rho^-$ of positive and negative edge dislocations are then given by
\begin{equation}
\frac{\partial \rho^+}{\partial t} = \partial_x(\rho^+ v) \quad,\quad \frac{\partial \rho^-}{\partial t} = - \partial_x(\rho^- v) \quad,
\end{equation}
or equivalently for the total dislocation density $\rho = \rho^+ + \rho^-$ and excess density $\kappa = \rho^+ - \rho^-$:
\begin{equation}
\frac{\partial \rho}{\partial t} = \partial_x(\kappa v) \quad,\quad \frac{\partial \kappa}{\partial t} = \partial_x(\rho v) \quad.
\end{equation}
The dislocation velocity $v$ is assumed in the form
\begin{equation}
v = \left\{\begin{array}{ll}
B (\tau_{\rm ext} + \tau_{\rm int}(\Br) - \tau_{\rm f}(\Br))\;,& \tau_{\rm ext} + \tau_{\rm int}(\Br) > \tau_{\rm f}(\Br)\;,\\
0\;,& \tau_{\rm ext} + \tau_{\rm int}(\Br) \le \tau_{\rm f}(\Br)\;.
\end{array}\right.
\end{equation}
Here the long-range internal stress $\tau_{\rm int}$ derives from the excess dislocation density $\kappa$ by
\begin{equation}
\tau_{\rm int}(\Br) = \int \kappa(\Br') \sigma_{xy}(\Br - \Br') {\rm d}^2 r' \;,\;
\end{equation}
where $\sigma_{xy} = G b g(\theta)/r$ is the $xy$ component of the edge dislocation stress field. The flow stress $\tau_{\rm f}(\Br)$ is assumed to relate to the total dislocation density in a non-local manner:
\begin{eqnarray}
\tau_{\rm f}(\Br) &=& \int \rho(\Br') \phi\left(\frac{\Br - \Br'}{\xi}\right) \sigma_{xy}(\Br - \Br') {\rm d}^2 r' \nonumber\\
&\approx&
\alpha G b \xi \rho(\Br) + \beta_x G b \xi^3 \partial_x^2 \rho(\Br) + \beta_y G b \xi^3 \partial_y^2 \rho(\Br)\;, 
\label{eq:flowstress2}
\end{eqnarray}
where $\phi$ is a correlation function of range $\xi$. In physical terms, $\xi$ characterizes the characteristic extension of the "jammed"  dislocation configurations (dipoles, multipoles, junctions) which control the flow stress. The non-dimensional coefficients $\alpha,\beta_x$ and $\beta_y$ of the long-wavelength expansion of the flow stress in Eq. (\ref{eq:flowstress2}) are given by
\begin{eqnarray}
\alpha &=& \int \phi(u,\theta) g(\theta) {\rm d} u {\rm d} \theta \;,\nonumber\\
\beta_x &=& \int u^2 \phi(u,\theta) g(\theta)\cos^2 \theta  {\rm d} u {\rm d} \theta\;,\nonumber\\
\beta_y &=& \int u^2 \phi(u,\theta) g(\theta)\sin^2 \theta  {\rm d} u {\rm d} \theta\;,
\end{eqnarray}
where $u = r/\xi$. 

We envisage a space- and time-independent reference state $\rho(\Br) = \rho_0, \kappa(\Br) = 0, \tau_{\rm ext} = \tau_0 + \tau_1$ where $\tau_1 = \alpha G b \xi \rho$ is the rate-independent part of the flow stress and the rate-dependent stress contribution $\tau_0$ corresponds to homogeneous plastic flow at rate $\dot{\gamma}_0 = \rho_0 b v_0 = \rho_0 b B \tau_0$. We analyze the evolution of space-dependent fluctuations $\delta \rho(x)$ and $\delta \kappa(x)$. For simplicity, we consider only fluctuations which are homogeneous in the $y$ direction, since such fluctuations do not give rise to long-range internal stresses. The temporal evolution of $\delta \rho(x)$ and $\delta \kappa(x)$ is in linear approximation given by 
\begin{eqnarray}
\label{eq:fluct}
\partial_t \delta \rho &=& B \tau_0 \partial_x \delta \kappa \;,\nonumber\\
\partial_t \delta \kappa &=& B \tau_0 \left[(1 - (\tau_1/\tau_0)(1+\eta)) \partial_x \delta \rho + (\beta_x/\alpha)(\tau_1/\tau_0) \xi^2 \partial_x^3 \delta \rho \right]\;.
\end{eqnarray}
where $\eta  = (\rho_0/\xi) (\partial \xi/\partial \rho|_{\rho_0})$. These evolution equations  are invariant under the scaling transformation (\ref{eq:trans}) if the characteristic length $\xi$ either scales in proportion with $1/\sqrt{\rho}$ or in proportion with $Gb/\tau_{\rm ext}$, or a mixture of both. In the former case, $\eta = 1/2$, and in the latter case, $\eta = 0$. 

\begin{figure}
\begin{center}
\includegraphics[width=0.7\textwidth]{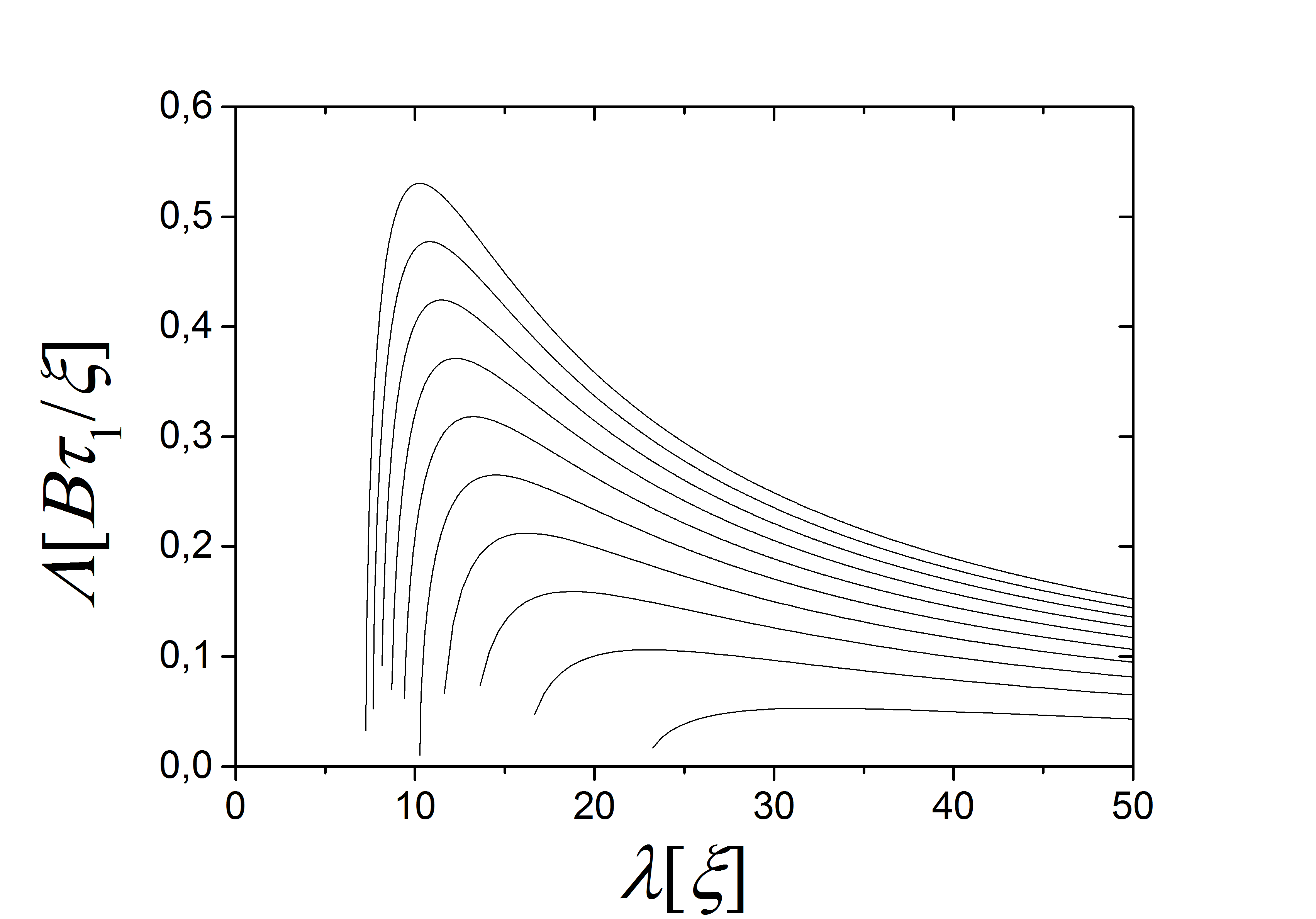}
\end{center}
\vspace*{-0.2cm}
\caption{Positive eigenvalue (amplification factor) as a function of pattern wavelength $\lambda$ for different values of the steady-state strain rate $\dot{\gamma}$; lowermost curve: $\dot{\gamma} = 0.9 \dot{\gamma}_{\rm c}$, uppermost curve: $\dot{\gamma} = 0.01 \dot{\gamma}_{\rm c}$; curves calculated for $\beta_x/\alpha = 2$.}
\label{fig:amplification}
\end{figure}

To investigate stability of the reference state, we make the Ansatz $\delta \rho(x,t) = \delta \rho(k) \exp(ikx) \exp(\lambda t)$ and $\delta \kappa(x,t) = \delta \kappa(k) \exp(ikx) \exp(\lambda t)$ which leads to the matrix equation
\begin{equation}
\displaystyle{
\Lambda \left[\begin{array}{l} \delta \rho \\ \delta \kappa \end{array} \right] = B \tau_0 
\left[\begin{array}{ll} 
0 & i k \\
i k [1 - (1+\eta)(\tau_1/\tau_0) - (\beta_x/\alpha)(\tau_1/\tau_0) \xi^2 k^2]
& 0 \end{array} \right]
\left[\begin{array}{l} \delta \rho \\ \delta \kappa \end{array} \right]}\;,
\end{equation}
with the eigenvalues
\begin{equation}
\Lambda = \pm (k B) \sqrt {\tau_1(1+\eta)- \tau_0 - (\beta_x/\alpha)\tau_1 \xi^2 k^2}\;.
\end{equation}
Positive real-valued eigenvalues exist if $\tau_1(1+\eta) > \tau_0$. In that case, a dominant wavelength (maximum of $\Lambda$) emerges at
$\lambda = 2 \pi \xi [(\alpha/2 \beta_x)(1 + \eta - \tau_0/\tau_1)]^{-1/2}$. In either of the two cases discussed above, $\xi = 1/\sqrt{\rho}$ or $\xi = Gb/\tau_{\rm ext}$, this wavelength is consistent with the similitude principle. A plot of the positive eigenvalue $\Lambda$ (the "amplification factor") as a function of wavelength and for different values of the steady-state strain rate is shown in Figure \ref{fig:amplification}. At low strain rate (quasi-static deformation), the dominant wavelength approaches a constant value of about 10$\xi$ (10 dislocation spacings).

It is interesting to have a closer look at the condition for symmetry breaking in this model, $\tau_1(1+\eta) > \tau_0$. This condition is tantamount to the requirement that the strain rate must not exceed the critical value $\dot{\gamma}_{\rm c} = B \tau_1(1+\eta) b \rho_0$ where the rate-dependent part of the flow stress, $\tau_0$, exceeds the rate-independent part $\tau_1$ by a factor $(1 + \eta)$. 
 In other words, the requirement for patterning is that the motion of dislocations must be mainly controlled by their interactions, rather than by the externally imposed driving stress. Patterning arises in this model from the fact that dislocation interactions impede the motion of dislocations -- hence, the dislocation velocity is reduced in regions of increased dislocation density which, in conjunction with the conserved nature of the transport dynamics, leads to further accumulation of dislocations in these regions. In the absence of plastic flow, no patterning is possible in this model since the dislocation arrangement remains frozen in the initial state. 

\section*{Appendix C: The stochastic model by H\"ahner and Zaiser}
\setcounter{equation}{0}
\renewcommand{\theequation}{C\arabic{equation}}

The model proposed by H\"ahner and Zaiser to describe the evolution of fractal dislocation microstructures is defined by two coupled stochastic evolution equations for densities of mobile and immobile dislocations, 
\begin{eqnarray}
\label{stoch1}
\partial_t \rho_{\rm m} = A \tau_{\rm ext} \langle \dot{\gamma} \rangle - B \sqrt{\rho} \dot{\gamma}\;,\\
\partial_t \rho_{\rm i} = (B-C) \sqrt{\rho} \dot{\gamma}\;.
\label{stoch2}
\end{eqnarray}
Here, the plastic strain rate is considered a randomly varying quantity with the statistical properties
\begin{eqnarray}
\langle \dot{\gamma} \rangle = \rho_{\rm m} b v\\
\langle \delta\dot{\gamma}(t)\delta\dot{\gamma}(t') \rangle = 2 \langle \dot{\gamma} \rangle^2 t_{\rm corr} \frac{\tau_{\rm int}}{S} \delta(t - t').
\end{eqnarray}
The correlation time $t_{\rm corr}$ is supposed to follow the relation $t_{\rm corr} \langle \dot{\gamma}\rangle = \rho_{\rm m} b L$ where $L$ is the dislocation mean glide path. The internal stress $\tau_{\rm int}$ is assumed to be approximately equal to the external stress. 

To analyze the scaling behavior of these equations under the transformations (\ref{eq:trans}) or (\ref{eq:trans1}) we first investigate how the  strain rate transforms.  Since $\rho_{\rm m,i} \to \lambda^{-2} \rho_{\rm m,i}$ and $v \to \lambda^{-n} v$ we find that $\langle \dot{\gamma} \rangle \to \lambda^{-2-n} \langle \dot{\gamma} \rangle$ (note that the transformation (\ref{eq:trans1}) can formally be considered by setting $n = -1$). The fluctuations transform in the same manner if the ratio $\tau_{\rm int}/S$ is a constant and the dislocation glide path transforms like $L \to \lambda L$. As discussed above, the former condition is fulfilled if the strain-rate sensitivity $S$ is controlled by overcoming of dislocation obstacles, and the latter if dislocation storage is controlled by dislocation interactions ($L \propto \rho^{-1/2}$). 

The left-hand sides of Eqs. (\ref{stoch1},\ref{stoch2}) transform under (\ref{eq:trans}) like $\partial_t \rho_{\rm m,i} \to \lambda^{-n-3} \partial_t \rho_{\rm m,i}$ and the model as a whole is invariant if all terms on the right-hand side transform in the same manner.  For the dislocation multiplication term this is the case if $A$ depends neither on stress nor on dislocation density, or if it depends on both variables only through a function of the invariant ratio $\tau_{\rm ext} \rho^{-1/2}$. The same is true for the constants $B$ and $C$ characterizing the dislocation storage terms in Eqs. (\ref{stoch1},\ref{stoch2}). If these conditions are fulfilled, the stochastic model can, even for time dependent stress $\tau_{\rm ext}$, be transformed to a non-dimensional form by setting
\begin{equation}
\rho_{\rm m,i} =  (A \tau_{\rm ext}^2/C)^2 \tilde{\rho}_{\rm m,i},\quad t =  A \tau_{\rm ext}^2/(C^2 \langle \dot{\gamma} \rangle) \tilde{t}, 
\label{stochscale}
\end{equation}
which automatically satisfies the scaling invariance requirements. 
With $\tilde{\rho} = \tilde{\rho}_{\rm m} + \tilde{\rho}_{\rm i}$, the resulting non-dimensional model is given by \cite{H"ahner99}
\begin{eqnarray}
\partial_{\tilde {t}} \tilde{\rho}_{\rm m} = 1- \tilde{\theta}\tilde{\rho}_{\rm m} - \frac{B}{C} \sqrt{\tilde{\rho}}(1 + \sigma \dot{w})  \;,\\
\label{stoch1a}
\partial_t \rho = 1 - \tilde{\theta}\tilde{\rho} -  \sqrt{\tilde{\rho}}(1 + \sigma \dot{w})\;.
\label{stoch2a}
\end{eqnarray}
where $\dot{w}$ is a standard correlated stochastic process, $\sigma$ the corresponding noise amplitude, and $\tilde{\theta} = (2A/C^2) \partial \tau_{\rm ext}/\partial \langle \gamma \rangle$ is a non-dimensional hardening coefficient which, in the regime of similitude scaling, is time independent (hardening stage II). Solutions of the Fokker-Planck equation corresponding to this stochastic model converge towards stationary solutions $p(\tilde{\rho})$; the corresponding solutions for the dimensional dislocation density describe a dislocation system whose stochastic signatures evolve with time only parametrically, through the scaling rules (\ref{stochscale}), and which thus in the course of hardening remains consistent with the similitude principle. Of course, deviations from this principle might be relevant -- e.g. if one includes an dynamic recovery term that is proportional to $\rho$ and which thus violates similitude scaling, the behavior at high stresses/dislocation densities will exhibit characteristic deviations from similitude as discussed in the main paper in the context of dislocation annihilation.

\section*{Appendix D: The Continuum Dislocation Dynamics (CDD) model by Hochrainer and co-workers}
\setcounter{equation}{0}
\renewcommand{\theequation}{D\arabic{equation}}

The model proposed by Hochrainer and co-workers \cite{Hochrainer2014_JMPS,Sandfeld2011_JMR,Hochrainer2013_MRS} describes the evolution of curved dislocations in a statistically averaged continuum model. We discuss this model here in order to illustrate how the scaling invariance principles we have formulated are reflected in a model which goes beyond standard dislocation density measures. 

In the model by Hochrainer and co-workers, the evolution of systems of dislocations  is described by a set of evolution equations for the total density $\rhot$, the vector of 'geometrically necessary dislocation' (GND) density $\Bkappa=[\kappa_1,\kappa_2]$, and a 'curvature density' $\qt$: 
\begin{eqnarray}
	\label{eq:drhotdt}
	\partial_t\rhot &=-\div(v\Bkappa^\perp)+v\qt\;,\\
	\label{eq:dkappadt}
	\partial_t\Bkappa &= -\curl(v\rhot\Bn)\;,\\
	\label{eq:dqtdt}
	\partial_t\qt&=-\div\left( -v\BQ^{(1)} + \BA^{(2)}\cdot \nabla v \right)\;,
\end{eqnarray}

where $\Bn$ denotes the slip plane normal and $\Bkappa^\perp=[\kappa_2,-\kappa_1]$. The evolution of these quantities depends on higher-order tensorial dislocation density measures, here the tensors $\BA^{(2)}$ and $\BQ^{(1)}$ in \eqref{eq:dqtdt}. These measures need to be related to the fields $\rhot,\Bkappa$ and $\qt$ through closure assumptions \cite{Monavari2013}, e.g. one may assume 
\begin{equation}\label{eq:closea1}
	{\BA}^{(2)} = \frac{1}{2}\left[ (\rhot +\kappa) {\Bl}_{\kappa} \otimes {\Bl}_{\kappa} + (\rhot -\kappa)	{\Bl}_{\kappa}^{\perp} \otimes {\Bl}_{\kappa}^{\perp}\right]
	\quad \textrm{and}\quad
  \BQ^{(1)} = -\Bkappa^\perp \frac{\qt}{\rhot}.
\end{equation}
Therein, ${\Bl}_{\kappa}^{\perp}$ is the unit vector perpendicular to ${\Bl}_{\kappa} = \Bkappa/\kappa$ which is a unit vector in the direction of the GNDs and $\kappa = |\Bkappa|$ is the scalar GND density.
Furthermore, the evolution of the plastic slip $\gamma$ is given by Orowan's law as
\begin{eqnarray} \label{eq:dgammadt}
\partial_t\gamma=\rhot b v.
\end{eqnarray}

To analyze the scaling behavior of these equations under the transformations (\ref{eq:trans}) or (\ref{eq:trans1}) we note that the dislocation velocity according to (\ref{eq:trans}) or (\ref{eq:trans1}) transforms like $v \to \lambda^{-n}v$ where (\ref{eq:trans1}) formally corresponds to the case $n = -1$. Thus, the strain rate transforms as $\partial_t\gamma\rightarrow\lambda^{-2-n}\partial_t\gamma$. The left-hand sides of Eqs. (\ref{eq:drhotdt}, \ref{eq:dkappadt}) transform under \eqref{eq:trans} as $\partial_t\rhot\rightarrow\lambda^{-n-3}\partial_t\rhot$ and $\partial_t\Bkappa\rightarrow\lambda^{-n-3}\partial_t\Bkappa$. For the right-hand sides to transform in the same manner, it is then necessary that the 'curvature density' must transform as $\qt \to \lambda^{-3} \qt$. This is consistent with the understanding of $\qt$ as a product 
of dislocation density and mean curvature, $\qt=\rhot k$ where $k$ has the dimension of a reciprocal curvature radius and thus transforms as $k \to \lambda^{-1}k$. The left-hand side of \eqref{eq:dqtdt} then obeys the transformation $\partial_t\qt\rightarrow\lambda^{-n-2}\partial_t\qt$. For the right-hand side to transform similarly, it is necessary that $\BQ^{(1)}$ transforms like $\qt$, and $\BA^{(2)}$ like $\kappa$ or $\rhot$. One easily ascertains that the closure equation (\ref{eq:closea1}) is consistent with this requirement. 

In order to scale the CDD model equations to a non-dimensional form we need to specify how the velocity $v$ depends on the dislocation fields. In principle, every dependence is admissible which leads to the correct scaling. For instance, we may write the dislocation velocity as a function of different interaction stresses, as e.g. a backstress $\taub$, a line tension contribution $\tault$ and a Taylor-type yield stress $\tauy$:
\begin{equation}
  \tau^{\rm b} = -{D\mu b}\frac{\nabla \cdot \Bkappa^\perp}{\rhot}\,,
  \qquad
  \tau^{\rm lt} = \frac{T}{b}  \frac{q^{\rm t}}{\rho^{\rm t}}\,,
  \qquad
  \tau^{\rm y} = \alpha G b \sqrt{\rho^{\rm t}}
  \label{eq:stresses}
\end{equation}
where $T\approx\mu b^2$ and the two non-dimensional parameters $D=0.6\ldots 1$ and $\alpha=0.2\ldots 0.4$. We can easily ascertain that all stresses in (\ref{eq:stresses}) have the correct scaling behavior. Assuming a linear relationship between stress and dislocation velocity ($n=1$) one gets
\begin{equation} \label{eq:v}
v = \left\{
\begin{array}{cc}
B(\tauext+\taub+\tault-\tauy) & \quad\textrm{if}\quad |\tauext+\taub+\tault|\geq|\tauy| \\ 
0                                       & \quad\textrm{else}
\end{array} 
\right.
,
\end{equation}
We may then define the following scaling relations between stresses $\tau$, densities $\rho$ and lengths $x$ and their dimensionless counterparts (indicated by the tilde)
\begin{eqnarray}\label{eq:scaling}
	\tau  =\alpha Gb\rho_0^{1/2} \,\tilde\tau,   \qquad
	\rho  =\rho_0 \,\tilde\rho , \qquad
	x     =D\rho_0^{-0.5}   \,\tilde x,
\end{eqnarray} 
where $\rho_0$ is the average initial dislocation density. By insertion into \eqref{eq:v} we can also derive scaling relations for the velocity $v$, the time $t$ and curvature density $\qt$ 
\begin{eqnarray}
v =  \frac{b^2}{B}\alpha G\sqrt{\rho_0} \,\tilde v,\qquad
t =  \frac{DB}{\alpha b^2G\rho_0}\,\tilde t,\qquad
\qt = \frac{\rho_0^{3/2}}{D} \,\tilde\qt.
\end{eqnarray}
Replacing all dimensional variables in \eqref{eq:v} by their scaled counterparts we obtain the non-dimensional velocity 
\begin{eqnarray}
\tilde v &=& \tilde\tauext 
   - \alpha^{-1} \frac{\tilde\nabla\cdot\tilde{\Bkappa}^\perp}{\tilde\rho}
   + (\alpha D)^{-1} \frac{\tilde q}{\tilde \rho}
   - \sqrt{\tilde\rho},
\end{eqnarray}
where $\tilde\nabla(\bullet)$ 
is the gradient operator w.r.t. the scaled coordinates. This equation only depends on two factors, $\alpha$ and $D$ which relate to dislocation pair correlation functions and thus characterize the mutual arrangement of dislocations; no material parameters occur which a posteriori justifies our choice in \eqref{eq:scaling}. Again, these equations automatically satisfy the scaling invariance requirements of similitude. The corresponding non-dimensional CDD evolution equations are then obtained from
\begin{eqnarray}
	\partial_t\tilde\rhot     = Q \, \partial_t\rhot, \qquad
	\partial_t\tilde{\Bkappa} = Q \, \partial_t\Bkappa, \qquad
	\partial_t\tilde\qt       = (Q\sqrt{\rho_0}/D) \; \partial_t\qt, \nonumber\\
	\partial_t\tilde{\gamma} = (QD/\sqrt{\rho_0}) \; \partial_t\gamma \qquad
	\textrm{where}\quad Q = \frac{\alpha b^2 G\rho_0^2}{BD}.
\end{eqnarray}
Analyzing the stability of these equations reveals patterning phenomena which are very similar to those discussed in Appendix B. A detailed discussion of these patterning phenomena will be presented elsewhere. Here, we only note that, again, the invariance of the fundamental equations under the scale transformations (\ref{eq:trans}) or (\ref{eq:trans1}) ensures that any dislocation patterns arising from these equations are consistent with the similitude principle.


\section*{References}
\begin{thebibliography}{99}
\bibitem{Kuhlmann1962} Kuhlmann-Wilsdorf D 1962 A new theory of work hardening {\it Trans. Met. Soc. AIME} {\bf 224} 1047-61
\bibitem{Raj1986} Raj SV and Pharr GM 1986 A compilation and analysis of data for the stress dependence of the subgrain size 
{\it Mater. Sci. Engng.} {\bf 81} 217-237
\bibitem{Sauzay11} Sauzay M and Kubin LP 2011 Scaling laws for dislocation microstructures in monotonic and cyclic deformation of fcc metals {\it Prog. Mater. Sci.} {\bf 56} 725�784
\bibitem{Rudolph05} Rudolph P, Frank-Rotsch C, Juda U and Kiessling FM 2005 Scaling of dislocation cells in GaAs crystals by global numeric simulation and their restraints by in situ control of stoichiometry {\it Mater. Sci. Engng. A} {\bf 400 - 401} 170-174 
\bibitem{Gomez06} Gomez-Garcia D, Devincre B and Kubin L 2006 Dislocation patterns and the similitude principle: 2.5d mesoscale simulations 
{\it Phys. Rev. Lett.} {\bf 96} 125503 
\bibitem{Zaiser01} Zaiser M, Miguel M-C and Groma I 2001 Statistical Dynamics of Dislocation Systems: The Influence of Dislocation-Dislocation Correlations {\it Phys. Rev. B}\ {\bf 64} 224102
\bibitem{Zaiser02} Zaiser M and Seeger A 2002 Long-Range Internal Stresses, Dislocation Patterning and Work Hardening in Crystal Plasticity, In: Dislocations in Solids Vol. 11, Eds. F.R.N. Nabarro, M. S. Duesbery and J. Hirth, North-Holland, Amsterdam, p. 1-99
\bibitem{Cottrell55} Cottrell AH and Stokes RJ 1955 Effects of temperature on the plastic properties of aluminium crystals {\it Proc. Roy. Soc. London A} {\bf 233} 17�34
\bibitem{Middleton92} Middleton AA 1992 Asymptotic uniqueness of the sliding state for charge-density wave {\it Phys. Rev. Lett.}\ {\bf 68}
670-673
\bibitem{Moretti04} Moretti P, Miguel MC, Zaiser M and Zapperi S 2004 Depinning transition of dislocation assemblies: Pileups and low-angle grain boundaries {\it Phys. Rev. B} {\bf 69} 214103
\bibitem{deWit60} de Wit R 1960  The continuum theory of stationary dislocations {\it Solid State Phys.} {\bf 10}
249-292
\bibitem{Madec03} Madec R, Devincre B, Kubin L, Hoc T and Rodney D 2003 The Role of Collinear Interaction in Dislocation-Induced Hardening {\it Science} {\bf 301} 1879-1882 
\bibitem{Madec02} Madec R, Devincre B and Kubin L 2002 From Dislocation Junctions to Forest Hardening {\it Phys. Rev. Lett} {\bf 89} 255508
\bibitem{Basinski79} Basinski SJ and Basinski ZS 1979 Plastic Deformation and Work Hardening, In: Dislocations in Solids Vol. 4, 
Eds. F.R.N. Nabarro and J. Hirth, North-Holland, Amsterdam, p. 261�362
\bibitem{VdG95} Van der Giessen E and Needleman A 1995 Discrete dislocation plasticity: a simple planar model {\it Modelling Simul. Mater. Sci. Eng.} {\bf 3} 689-735
\bibitem{Brown64} Brown LM 1964 The self-stress of dislocations and the shape of extended nodes {\it Phil. Mag.} {\bf 10} 441-466
\bibitem{Gavazza76} Gavazza SD and Barnett DM The self-force on a planar dislocation loop in an anisotropic linear-elastic
medium {\it J. Mech. Phys. Solids} {\bf 24} 171 - 185
\bibitem{Essmann79} Essmann U and Mughrabi H 1979 Annihilation of dislocations during tensile and cyclic deformation and limits of dislocation densities {\it Phil. Mag. A} {\bf 40} 731 - 756
\bibitem{Miguel02} Miguel MC, Vespignani A, Zaiser M and Zapperi S 2002 Dislocation jamming and Andrade creep {\it Phys. Rev. Letters} {\bf 89} 165501
\bibitem{Zbib98} Zbib HM, Rhee M and Hirth JP 1998 On plastic deformation and the dynamics of 3D dislocations {\it Int. J. Mech. Sci.} {\bf 40}
113�127
\bibitem{Cai06} Cai W, Arsenlis A, Weinberger CR and Bulatov VV 2006 A non-singular continuum theory of dislocations {\it J. Mech. Phys. Solids}
{\bf 54} 561 � 587
\bibitem{Brown02} Brown LM 2002 A dipole model for the cross-slip of screw dislocations in fcc metals {\it Phil. Mag. A} {\bf 82} 1691-1711
\bibitem{Paus13} Paus P, Kratochvil J and Benes M 2013 A dislocation dynamics analysis of the critical cross-slip annihilation distance and the cyclic saturation stress in fcc single crystals at different temperatures {\it Acta Mater. 61} 7917�7923
\bibitem{Thornton62} Thornton PR, Mitchell TE and Hirsch PB 1962 The dependence of cross-slip on stacking-fault energy in face-centred cubic metals and alloys {\it Phil. Mag.} {\bf 7} 1349-1369
\bibitem{Kubin92} Kubin LP, Canova G, Condat M, Devincre B, Pontikis V and Br�chet Y 1992 Dislocation microstructures and plastic flow: a 3D simulation. {\it Solid State Phenomena} {\bf 23} 455-472
\bibitem{Rhee98} Rhee M, Zbib HM, Hirth JP, Huang H. and De la Rubia T 1998 Models for long-/short-range interactions and cross slip in 3D dislocation simulation of BCC single crystals {\it Modell. Sim. Mater. Sci. Engng.} {\bf 6} 467-492
\bibitem{Bonneville88} Bonneville J, Escaig B and Martin JL 1988 A study of cross-slip activation parameters in pure copper  {\it Acta Metall.}
{\bf 36} 1989-2002 
\bibitem{Bulatov98} Bulatov, V., Abraham, F. F., Kubin, L., Devincre, B. and Yip, S. 1998 Connecting atomistic and mesoscale simulations of crystal plasticity {\it Nature} {\bf 391} 669-672
\bibitem{VanderGiessen95} Van der Giessen E and Needleman A 1995 Discrete dislocation plasticity: a simple planar model {\it Modelling Simul. Mater. Sci. Eng.} {\bf 3} 689-735
\bibitem{Benzerga04} Benzerga AA, Brechet Y, Needleman A and Van der Giessen E 2004 Incorporating three-dimensional mechanisms into two-dimensional dislocation dynamics {\it Modell. Sim. Mater. Sci. Engng.} {\bf 12}, 159-196
\bibitem{Holt70} Holt DL 1970 Dislocation cell formation in metals {\it J. Appl. Phys.} {\bf 41} 3197-3201
\bibitem{Ananthakrishna81} Ananthakrishna G and Sahoo D 1981 A model based on nonlinear oscillations to explain jumps on creep curves {\it J. Phys. D: Applied Physics} {\bf 14} 2081-2090
\bibitem{Walgraef85} Walgraef D and Aifantis EC 1985 Dislocation patterning in fatigued metals as a result of dynamical instabilities {\it J. Appl. Phys.} {\bf 58} 688-691
\bibitem{Groma97} Groma I 1997 Link between the microscopic and mesoscopic length-scale description of the collective behavior of dislocations {\it Phys. Rev. B} {\bf 56} 5807-5813
\bibitem{Groma03} Groma I, Csikor FF and Zaiser M 2003 Spatial correlations and higher-order gradient terms in a continuum description of dislocation dynamics {\it Acta Mater.} {\bf 51} 1271-1281
\bibitem{H"ahner96} H\"ahner P 1996 A theory of dislocation cell formation based on stochastic dislocation dynamics {\it Acta Mater.} {\bf 44}
2345-2352
\bibitem{H"ahner99} H\"ahner P and Zaiser M 1999 Dislocation dynamics and work hardening of fractal dislocation cell structures {\it Mater. Sci. Engng. A} {\bf 272} 443-454
\bibitem{Aifantis84} Aifantis EC 1984 On the microstructural origin of certain inelastic models {\it J. Engng. Mater. Technol.} {\bf 106} 326-330.
\bibitem{Hochrainer2014_JMPS} Hochrainer T, Sandfeld S, Zaiser M and Gumbsch P 2014
Continuum dislocation dynamics: Towards a physical theory of crystal plasticity
{\it J. Mech. Phys. Solids} {\bf 63} 167-178 
\bibitem{Devincre08} Devincre B, Hoc T and Kubin L 2008 Dislocation mean free paths and strain hardening of crystals {\it Science} {\bf 320} 1745-1748
\bibitem{Zhou12} Zhou C, Reichhardt C, Olson Reichhardt CJ and Beyerlein IJ 2012 Dynamic phases, pinning, and pattern formation for driven dislocation assemblies, arXiv cond-mat.mtrl.sci: 1207.6657
\bibitem{Tsekenis11} Tsekenis G, Goldenfeld N and Dahmen KA 2011 Dislocations jam at any density {\it Phys. Rev. Lett.} {\bf 106} 105501
\bibitem{Zaiser06} Zaiser M 2006 Scale invariance in plastic flow of crystalline solids {\it Adv. Phys.} {\bf 55} 185-245
\bibitem{Zaiser99} Zaiser M and H\"ahner P 1999 The flow stress of fractal dislocation arrangements {\it Mater. Sci. Engng. A} {\bf 270} 299-307
\bibitem{Tang07} Tang H, Schwarz KW and Espinosa HD 2007 Dislocation escape-related size effects in single-crystal micropillars under uniaxial compression {\it Acta Mater.} {\bf 55} 1607-1616
\bibitem{Greer11} Greer JR and De Hosson JTM 2011 Plasticity in small-sized metallic systems: Intrinsic versus extrinsic size effect {\it Progr. Mat. Sci.} {\bf 56} 654-724
\bibitem{Uchic09} Uchic MD, Shade PA and Dimiduk DM 2009 Plasticity of micrometer-scale single crystals in compression {\it Ann. Rev. Mater. Res.} {\bf 39} 361-386
\bibitem{Groma06} Groma I, Gyorgyi G and Kocsis B 2006 Debye screening of dislocations {\it Phys. Rev. Letters} {\bf  96} 165503
\bibitem{Sandfeld2011_JMR} Sandfeld S, Hochrainer T, Zaiser M and Gumbsch P 2011
Continuum modeling of dislocation plasticity: Theory, numerical
implementation, and validation by discrete dislocation simulations
{\it J. Mater. Res.} {\bf 26} 623-632
\bibitem{Hochrainer2013_MRS} Hochrainer T 2014
Continuum dislocation dynamics based on the second order alignment tensor
{\it Mater. Res. Soc. Symp. Proc.} {\bf 1651} DOI: 10.1557/opl.2014.53
\bibitem{Monavari2013} Monavari M, Zaiser M and Sandfeld S 2014 
Comparison of closure approximations for continuous dislocation dynamics
{\it Mater. Res. Soc. Symp. Proc.} {\bf 1651}  DOI: 10.1557/opl.2014.62

\end{thebibliography}
\end{document}